# Information -sharing, adaptive epigenetics and human longevity

Author:

Marios Kyriazis,

ELPIs Foundation for Indefinite Lifespans,

email: drmarios@live.it

Abstract

Emerging empirical and theoretical thinking about human aging places considerable value upon the role of the environment as a major factor which can promote prolonged healthy longevity. Our contemporary, 'information-rich' environment is taken to mean not merely the actual physical surroundings of a person but it is also considered in a more abstract sense, to denote cultural, societal and technological influences. Our modern environment is far from being static or stable. In fact, it is continually changing in an exponential manner, necessitating constant adaptive responses on behalf of our developmental and evolutionary mechanisms. In this paper, I attempt to describe how a continual, balanced and meaningful exposure to a stimulating environment, including exposure to 'information-that-requires-action' (but NOT trivial information), has direct or indirect repercussions on epigenetic mechanisms which may then act to prolong healthy longevity. Information gained from our environment acts as a hormetic stimulus which up-regulates biological responses and feedback loops, eventually leading to improved repair of age-related damage. The consequence of this biological information-processing mechanism may influence resource allocation and redress the imbalance between somatic cell versus germ-line cell repairs. This can eventually have evolutionary consequences resulting in the drastic reduction of age-related disease and degeneration.

Introduction

Adaptive phenotypic plasticity is an important factor in human evolution. There is considerable evidence suggesting that the environment influences the plasticity of the phenotype [1] and that there exist precise, if hitherto not very well studied, mechanisms, which respond to stimulating changes in the environment, in an attempt to adapt to the stimulus [2]. This adaptation to a stimulus is very relevant in the study of age-related pathology and in the evolution of aging in general. Increasingly, research shows that the environment is crucial in determining prolonged healthy longevity [3]. In human terms, the 'environment' is an abstract notion of an amalgam of physical and virtual surroundings, interactions with modern society, and techno-cultural elements. It is, in other words, a highly 'information-rich' milieu. It is worth noting at this point that I define information as *'a meaningful set of data or patterns which influence the formation or transformation of other data or patterns, in order to reduce uncertainty and help achieve a goal'* [4] (FIGURE 1).

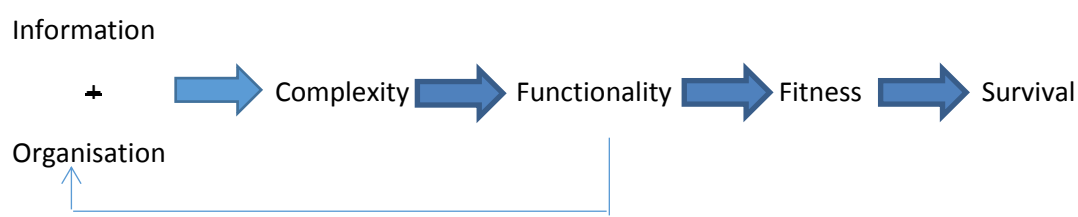

**Figure 1**. Information (plus organisation) increases complexity and this increases functionality. This improves fitness and thus survival. Any increase in internal fitness requires the formation of new links and the strengthening of the interconnectedness between its nodes i.e. increased complexity, thus increased fitness and increased survival.

However, as our environment is constantly changing, (and recently, changing at an ever-accelerating manner [5]) we need to quickly adapt to these changes in order to be able to survive. This may depend on artificial selection (through technology and, for example, exposure to meaningful information) rather the slower Darwinian natural selection processes [6]. Epigenetic factors mediate between the genome and the environment, and play an important role in this respect, allowing for rapid changes in protein expression which may be applied almost immediately when needed.

There is an increasing body of research which highlights the role of epigenetic changes in age-related degeneration [7]. Epigenetic changes such as chromatin remodelling and DNA methylation can modify the phenotype and play a modulating role in senescence and age-disease. One of the reasons for this, is based on the fact that epigenetic processes can regulate mechanisms necessary for development [8]. The process of aging has been explained by several authors as a continual and excessive operation of basic developmental, which eventually lead to time-related malfunction [9, 10], therefore a corrective regulation of this defective process could lead to the control of at least some aging mechanisms. In addition, it is known that there could be 'molecular breaks' which restrict the plasticity of adult cells, in particular neurons [11]. The use of epigenetic interventions may lead to stabilisation and delayed expression of these break factors and thus allow for easier and quick plasticity of the cell. The fact that several of these factors are found in the extracellular space makes it more likely that epigenetic control would be more effective compared to genetic intracellular mechanisms.

Hormesis and stress

Several environmental factors are closely linked to the epigenetic regulation of gene expression. This epigenetic regulation may lead to adaptive changes, which increase the likelihood of successful survival within a particular niche. The epigenetic changes can be predictable, can occur both during early and late developmental stages, and can result in improvement of the functional ability of the organism [12]. It has been shown that environmental stress induces epigenetic changes and results in an adaptive response to the stress in question [13]. The question arises: "what is stress?" In this context stress is taken to mean *exposure to a stimulus which disrupts homeostasis and necessitates a reaction in order for the organism to adapt to this stimulus*.

Stress (or more colloquially 'positive stress') can be of hormetic origin i.e. a dose-response phenomenon (with low-dose stimulation and high-dose inhibition) [14-17]. Hormesis is considered as a process of continual novelty, which affects basic biological parameters [18]. Vaiserman [19] considers that a mild stress-induced hormetic response involves mechanisms similar to the mechanisms that underlie developmental epigenetic adaptations, with resulting life-span prolonging properties. Examples of positive hormetic stress include calorie restriction, physical exercise, and certain chemical compounds including free radicals and mild toxins. In addition, there could be other types of hormetic stimulation such as cultural, societal and psychological factors, all of which interact at multiple levels. It can be argued that hormetic stress, as the type encountered in our everyday highly technological, intensely information-sharing environment, may cause direct epigenetic changes, which facilitate our adaptation to this environment. But, by scrutinising these epigenetic changes more closely, one may find that there are similarities with some of the mechanisms that repair age-related damage. This may be due to rapid up-regulation of genes responsible for adaptive response, which can have an effect on healthy life extension [20]. It has to be highlighted however that excessive stress is a well-known determinant for disease.

Although epigenetic adaptations to a changing environment can occur during early stages of development, it is also possible to experience epigenetic plasticity in later life stages [21]. This is a promising fact, which supports the view that hormetic stressors can up-regulate healthy longevity mechanisms even if applied later in life. Although the search for an ideal type of mild positive (hormetic) stress still continues, it makes sense to consider stimuli which are directly relevant to our current technological and information-rich environment; stresses that must lead to adaptation. For instance, exposure to a relentless information-sharing environment, such as meaningful online hyper-connectedness, which acts as a kind of hormetic stimulus would lead to an adaptive response which may up-regulate several health parameters [22] as discussed below. Once again it must be emphasized that the issue here concerns meaningful stimulation containing information that requires action, and not trivial or meaningless information.

Resource allocation

How can an information-rich environment affect biological parameters and lead to a reduced rate of senescence? It has been suggested that although efficient repair mechanisms are present in somatic cells these are traded-off in germ-line cells against optimal and successful survival (of the species). Heininger [23] quotes:

> *Evolution through natural selection can be described as driven by a perpetual conflict of individuals competing for limited resources.* **Somatic death is a germ cell-triggered event** (emphasis mine*) and has been established as evolutionary-fixed default state following asymmetric reproduction in a world of finite resources. Aging, on the other hand, is the stress resistance-dependent phenotype of the somatic resilience that counteracts the germ cell-inflicted death pathway. Thus, aging is a survival response and, in contrast to current beliefs, is antagonistically linked to death that is not imposed by group selection but enforced upon*

> *the soma by the selfish genes of the "enemy within" (*i.e. the germ-line cells, comment mine*). Environmental conditions shape the trade-off solutions as compromise between the conflicting germ–soma interests. Mechanistically, the neuroendocrine system, particularly those components that control energy balance, reproduction and stress responses, orchestrate these events.*

Therefore, the concept of aging can be viewed, at least partly, as a conflict between germ-line cells and somatic cells, whereby immortalisation factors are preferentially diverted towards germ-line cells, in order to assure the survival of the species, instead of being used by the somatic cells [24]. This results in a continual effective repair of the germ cells, which exhibit robust anti-senescence traits, at the expense of the somatic cells which eventually become overwhelmed by damage. This maintenance gap is a matter of both *resources required* and *resources allocated*, that is to say that if resources are constantly required, these will be allocated in order to assure long term survival, despite the short-term evolutionary cost [25]. The main existing evolutionary reason for this is the need to assure the survival of the species, which can currently be achieved much more energy-efficiently through immortalisation of germ line cells.

I have suggested that it may be possible to re-balance the germ-soma conflict using continual exposure to insightful information, a mechanism which depends on epigenetic modification and is in accordance with basic evolutionary requirements [26]. This is a typical example of how environmental effects (i.e. hyper-connection and a continual exposure to digital technology for example) can have a direct impact upon age-regulating mechanisms via epigenetic instead of genetic changes. It is possible to examine further the role of epigenetic mechanisms in relation to the shifting of resources from germ-line to somatic repairs. For instance, it is known that there exist non-autonomous contributions of somatic cells to germ line cells, leading to germ-line immortalisation [27]. These contributions aid germ-line survival and immortalisation. Such strategies may also be present in somatic cells, but are significantly down-regulated [28]. The aim here is to examine how this rejuvenation process can be driven to operate in somatic cells [29]. It is important to highlight that certain mechanisms of germ line rejuvenation could be dependent upon epigenetic modifications and factors that regulate transcription [30]. Thus it is also plausible that epigenetic co-ordination may have certain rejuvenating effects upon somatic cells although, clearly, more research is needed in this area. It is also worth remembering that not all epigenetic regulation may lead to positive results. For example, increasing methylation may down-regulate a deleterious signalling pathway but it may also down- regulate tumour suppressor genes increasing the risk of cancer [31, 32]. This also needs further research.

This conflict for survival between somatic and germ-line cells can be conceivably reversed if the survival of somatic cells becomes a priority over the survival of germ-line cells. The mechanism for this is likely to involve hormetic epigenetic influences of the environment upon the soma. The effectiveness of this process may be time-dependent, i.e. may depend on how early in life the organism is subjected to the 'stressor'. Some evidence exists that if hormetic, epigenetic stressors are applied early in life, these will have a more pronounced and lasting effect [33].

Flatt et al. [34] have shown that changes in the environment have a direct effect upon the regulation of somatic maintenance and they suggest that the same genotype can express either of two alternative phenotypes, depending on the particular environment. The example supports in principle the suggestion that suitable manipulation of the environment (via exposure to relevant digital information, for example) can have effects on the human epigenome and modulate aging. Calvanese et al. [35] suggest that the role of epigenetic modifications during aging is defined by the need for these specific epigenetic changes to exist, and, that these changes must be functionally associated with the aged phenotype.

The impact of information upon the cell

As mentioned above, information is meaningful sets of data which influence the formation or transformation of other data in order to reduce uncertainty (and thus entropy) and achieve a goal. In the practical sense, I assume that Shannon's information entropy [36] is similar to thermodynamical entropy, and the discussion is based upon this general model [37].

Most authorities agree that Environmental Enrichment (EE) (i.e. a stimulating and changing 'information-rich' environment) and hormesis have an effect on several parameters, such as an up-regulation of Brain Derived Neurotrophic Factor. In this particular case, it was also shown that an enriched (information-rich) environment <u>early in life</u> results in highly up-regulated histone acetylation at the BDNF gene <u>at adulthood</u>, which happens quite fast after a stimulating exposure [38]. With regards to BDNF, it is also known that its concentration is raised by exercise [39]. This has direct relevance here, and I quote from my own paper [26]:

> *The suggestion that a sustained exposure to actionable information forcibly induces a change in basic hitherto stable evolutionary processes is not an entirely speculative one. It has been proposed that increased aerobic activity (thus in this case physical, instead of cognitive, effort) had a direct evolutionary impact on the human brain [40]. The increased and sustained aerobic activity of hunter-gatherers resulted in the up-regulation of peripheral BDNF and increased its concentration to such a degree that it eventually crossed the blood-brain barrier and directly influenced the function and survival of the neurons. The plasma concentrations of other factors that are normally found in the periphery (i.e. not in the brain) such as insulin-like growth factor 1 (IGF-1) and vascular endothelial growth factor (VEGF) all of which increase during physical activity, were elevated by the sustained physical effort and, having crossed the blood-brain barrier, reached the brain resulting in improved neurogenesis. Thus, it is hypothesized that other sustained activities, in this case cognitive stimulation, may have physical effects which cause a transition from one evolutionary stage to another.*

It is known that several epigenetic mechanisms, including methylation of DNA and chromatin remodelling mediate gene regulation in neurons and thus have an important effect in regulating the development of the nervous system. These epigenetic modelling effects shape neural structure and function, and depend upon external environmental clues. Epigenetic gene regulation is important in regulating synaptic plasticity, neural behaviour and high order cognitive functions such as memory, learning and problem-solving abilities [41].

In order to gain better insights into how information may affect epigenetic regulation, it is useful to explore in some detail the process of information manipulation by the cell. In this case, a relevant example to consider is visual information. When new information is captured by retinal cells there is photonic recycling on the cell surface [42]. Ligand-receptor associations alter the conformation of the extracellular portion of intramembranous proteins and this change is transmitted to the cytoplasm by the trans-membranous helical segments by non-linear vibrations of proteins [42]. In addition, photon-photon interactions induce molecular vibrations responsible for bio-amplification of weak signals described by:

$mc^2$=BvLq

where *m* is the mass of the molecule, *c* is the velocity of the electromagnetic field, B is the magnetic flux density, v is the velocity of the carrier in which the particle exists, L is its dimension, and q is a unit charge.

When the signal is transmitted to the appropriate neuronal areas (in his case in the occipital cortex) it effects a neuronal response in terms of mainly epigenetic modulation such as DNA acetylation. The steps leading from the acquisition to the assimilation of the information require energy, and this energy needs to be diverted away from other processes in order to be made available to the information-manipulation process. These adaptation mechanisms depend partly on genetic function but mainly on epigenetic factors, which are able to act faster than a genetically-originated reaction, and provide more flexibility for optimising performance. It is known that epigenetically-dependent adaptation to visual stimulation can be very rapid [43]. Thus, the environment, through epigenetic regulation affects brain function, and it follows that if environmental stimulation is of a certain magnitude and quality, this will have direct effects on cognition. Cognition and higher mental functions are essential in our current technological, high information-dependant environment, and those who are able to adapt to these challenges are more likely to survive within this milieu.

Recent research has shed further light in some of the mechanisms of the effects of the environment upon epigenetic regulation. For instance, it is known that genetic elements such as microRNAs (miRNAs) are effective epigenetic regulators of a genomic response to environmental stimuli. Environmental agents such as $O_3$ exposure can disrupt miRNA expression profiles and interfere with inflammation regulation [44]. MicroRNAs contribute to genome plasticity and evolution, and are essential in controlling the expression of genes which may interfere with the aging process. However, further study is needed in this respect, as many effects of miRNA regulation are still unknown [45]. What is known is that the number and function of miRNAs is significantly up-regulated in humans compared to other animals such as chimpanzee, macaque, mouse, rat or dog [46]. This may indicate that humans may have developed unique evolutionary genetic elements that can facilitate a better repair of damage, exactly because of the need to adapt quickly to an information-rich technological environment [47], an evolutionary situation that is not relevant in other non-human animals.

Finally, it is worth mentioning that a chronically unpredictable environment (i.e. as our constantly changing society and techno-culture) can induce phenotypic changes in the parents, which can be reflected in the unexposed offspring, suggesting that at least some of the epigenetic changes can be heritable. This adds robustness to the notion suggesting that epigenetic changes are important in evolution, as these are more permanent and can be subjected to evolutionary selection. Also, repeated stress early in life can protect against more sustained stress later in life [48] and social factors can play a role in epigenetic regulation the effects of which not only persist throughout life, but can also be transmitted to the offspring via epigenetic inheritance [49]. In addition, there are specific epigenetic profiles associated with longevity [50] which can be transmitted to the offspring. Gentilini et al. [51] have shown that centenarians have certain specific DNA methylation characteristics (such as those found in nucleodite biosynthesis, and control of signal transmission) which are also found in their offspring, suggesting that these characteristics are heritable.

Conclusion

Epigenetic regulation mediates adaptation to the environment, and when this environment is of a specific type (an information-sharing, fast changing techno-cultural setting) then fast adaptation to this may have an impact upon life-extending biological processes. This epigenetic regulation, together with genetic and cell-specific factors can influence longevity and have been studied within a non-reductionist, integrative framework of aging [52]. The reductionist view that aging can be manipulated by simple biomedical repairs is unlikely to lead to any appreciable practical results that can be used in order to diminish the impact of age-related degeneration. Instead, a wider approach more likely to succeed would be to study the role of complexity, regulation and adaptation, and harness the function of fundamental processes such as the role of emergence and the 'direction' of human evolution [53]. This world-view has been encapsulated in philosophical reflections which transcend time and place, such as *"Tempora mutantur, nos et mutamur in illis"* ("Times change, and we change with them"). As our environment, society and culture change, so must we, in order to adapt to the change and increase our fitness and chance of survival [54]. One way we can effect a positive change is to increase exposure to meaningful technological information that requires action [55], and exploit the nature of relatively fast epigenetic mechanisms which, bound by evolutionary constraints, must result in suitable beneficial adaptations, and thus extended survival.